\documentclass[12pt]{article}
\pdfoutput=1 
\usepackage{jheppub} 
\usepackage{physics}
\usepackage{simpler-wick}

\usepackage[utf8]{inputenc}
\usepackage[T1]{fontenc} 
\usepackage{color}
\usepackage{verbatim}
\usepackage{amsmath}
\usepackage{array}
\usepackage[normalem]{ulem}

\pdfoutput=1 

\usepackage{jheppub}
\usepackage{appendix}

\newcommand{\be}{\begin{equation}}
\newcommand{\ee}{\end{equation}}
\newcommand{\bea}{\begin{eqnarray}}
\newcommand{\eea}{\end{eqnarray}}
\newcommand{\bi}{\begin{itemize}}
\newcommand{\ei}{\end{itemize}}

\def\ba#1\ea{\begin{align}#1\end{align}}
\def\bg#1\eg{\begin{gather}#1\end{gather}}
\def\bm#1\em{\begin{multline}#1\end{multline}}
\def\bmd#1\emd{\begin{multlined}#1\end{multlined}}
\def\a{\alpha}
\def\b{\beta}
\def\c{\chi}

\def\d{\delta}

\def\G{\Gamma}

\def\m{\mu}
\def\n{\nu}
\def\p{\phi}

\def\r{\rho}
\def\s{\sigma}

\def\t{\tau}

\newcommand{\fr}{\frac}

\newcommand{\pa}{\partial}

\newcommand{\wtd}{\widetilde}

\newcommand{\nn}{\nonumber}
\newcommand{\qu}{\quad}

\renewcommand{\(}{\left(}
\renewcommand{\)}{\right)}
\renewcommand{\[}{\left[}
\renewcommand{\]}{\right]}

\title{\boldmath Orbit Averaging Coherent States: Holographic Three-Point Functions of AdS Giant Gravitons}

\author{Adolfo Holguin,}
\author{Wayne W. Weng}

\affiliation{Department of Physics, University of California, Santa Barbara, CA 93106, USA}

\emailAdd{adolfoholguin@physics.ucsb.edu}
\emailAdd{wweng@physics.ucsb.edu}

\abstract{We study correlation functions of two AdS giant gravitons in AdS$_5\times S^5$ and a BPS supergravity mode using holography. In the gauge theory these are described by BPS correlators of Schur polynomials of fully-symmetric representations and a single trace operator. We find full agreement between the semiclassical gravity and gauge theory computations at large $N$, for both diagonal and off-diagonal structure constants. Our analysis in $\mathcal{N}=4$ SYM provides a simpler derivation to the results in the literature, and it can be readily generalized to operators describing bound states of AdS giant gravitons as well as bubbling geometries.}

\begin{document} 
\flushbottom

\maketitle
\section{Introduction}
The AdS/CFT correspondence provides in principle a way of addressing interesting questions in simple theories of quantum gravity \cite{Maldacena:1997re}. However the usual lore states that this is a weak/strong duality; objects that behave classically in gravity are described by complicated states in a strongly coupled conformal field theory. Fortunately this is not the case, as protected operators with large dimensions can and do behave semiclassically on both sides of the duality.

One of the simplest example of such an object is a half-BPS determinant operator in $\mathcal{N}=4$ SYM
\begin{equation}
    \mathcal{D}(x,\xi)=\det\left( \mathbf{1}\xi - Z(x)\right)=\int d\bar{\chi} d\chi \exp\left(-\bar{\chi}\left[\xi-Z(x)\right]\chi\right),
\end{equation}
whose dual description is a wrapped D3-brane inside of $S^5$, sitting at the origin of AdS \cite{Corley:2001zk}. The fact that these operators describe localized probes of AdS$_5\times S^5$ makes them ideal probes for bulk locality. The main obstacle to dealing with such objects on the gauge theory lies in the sheer combinatorial complexity of summing large numbers of planar graphs. Recently, this problem was revisited by using saddle-point methods to systematically resum these non-planar contributions \cite{Jiang:2019xdz, Chen:2019gsb}. This allows for an efficient computation of simple correlators involving determinant operators in the large $N$ limit. As an application, the authors of \cite{Yang:2021kot} studied the three-point function of a BPS single trace operator and two determinants and found a remarkable agreement with the \textit{orbit average} of the holographic computation of \cite{Bissi:2011dc}. Holographic three-point functions of giant gravitons have been studied extensively in the literature \cite{Berenstein:1998ij, Lee:1998bxa, Roiban:2010fe, Hernandez:2010tg, Ryang:2010bn, Georgiou:2010an, Zarembo:2010rr, Russo:2010bt, Costa:2010rz, Bak:2011yy, Bissi:2011dc, Caputa:2012yj,Lin:2012ey, Hirano:2012vz, Kristjansen:2015gpa, Jiang:2019zig}, with some discrepancies and ambiguities appearing between the holographic and gauge theoretic computations in the case of off-diagonal extremal correlators and for AdS giant gravitons. 

In \cite{Chen:2019gsb, Yang:2021kot}, similar techniques were introduced for studying fully-symmetric Schur polynomial operators:
\begin{equation}
    \mathcal{S}(x, \xi) = \int_{\mathbb{C}^N} d \bar{\varphi} d\varphi\exp\left(-\bar{\varphi}\left[\xi-Z(x)\right]\varphi\right)= \frac{1}{\det\left( \mathbf{1}\;\xi - Z(x)\right)}.
\end{equation}
Formally, this object is a generating function for BPS operators transforming in fully-symmetric representations of $U(N)$, which describe giant gravitons extended along the AdS$_5$ directions.  Despite the similarities between the techniques developed for determinant operators, these generating functions have an important distinction in that they do not correspond to simple semiclassical states. In fact, these generating function create a non-physical state of infinite norm in $\mathcal{N}=4$ SYM. The symmetry between sphere and AdS giants can be restored by considering BPS coherent states in the gauge theory \cite{Berenstein:2022srd}. These are given by a group averages of the exponential of one of the complex scalar fields. 

The goal of this paper is to extend the analysis in \cite{Yang:2021kot} to the case of AdS giant gravitons and to further clarify some technical aspects of their computation. Our analysis essentially mirrors \cite{Bissi:2011dc} but the set-up and results are different. After performing an orbit average of the semiclassical one-point functions of a BPS supergravity mode, we find precise agreement with the gauge theory computation of a BPS three-point function involving two heavy symmetric Schur polynomials and a single trace operator. Despite the fact that the intermediate steps in the computation are rather different from the case of sphere giants we find that the final results are related by a simple analytic continuation.  Our derivation of the structure constant in $\mathcal{N}=4$ SYM is new, and also involves a sort of orbit average,  although its relation to the one in holography is unclear. As we will explain, our methods have straightforward generalizations to the case of correlators of more general Schur polynomials, although we leave the details of this analysis for future work.

The paper is structured as follows. In section \ref{2} we review the orbit average method and how it applies to holographic correlation functions. Then, we review the coherent state techniques necessary for the large $N$ analysis in the field theory in section \ref{3}. In section \ref{4} we turn to computation of the structure constant of two BPS fully-symmetric Schur polynomials and a single trace BPS operator. We provide an exact integral formula for the generating function for these structure constants, which we then evaluate via the saddle-point approximation. In section \ref{5} we compute diagonal and off-diagonal structure constants in the dual supergravity following \cite{Yang:2021kot}, finding an exact matching with the gauge theory result. Finally we comment on possible future directions.

\section{Review of Orbit Average}\label{2}
We begin by giving a brief review of the semiclassical techniques found in \cite{Monin:2016jmo, Bajnok:2014sza} and \cite{Yang:2021kot}, known as the orbit average method.

The idea is as follows: consider a quantum mechanical system whose action $S\[X\]$ is invariant under some global symmetry $G$. In general, the eigenstates of the Hamiltonian will not be invariant under this global symmetry, but will rather transform in some representation of $G$ labelled by a set of charges $\{J_i\}$. One is usually interested in computing correlation functions of operators in backgrounds with non-zero charges
\bea
C_{JJ'\mathcal{O}_L} = \langle J'| \mathcal{O}_L(t=0) | J \rangle,
\eea
where we can think of the states $\ket{J}, \ket{J'}$ as being created by the insertion of operators with large charges, $J, J' \gg 1$. In the WKB approximation, this quantity can be computed by a path integral with the corresponding classical action evaluated on solutions to the equations of motion:
\begin{equation}
    \langle J'| \mathcal{O}_L(t=0) | J \rangle\sim e^{i S[X^*]} \mathcal{O}[X^*],
\end{equation}
where $\mathcal{O}_L[X] \equiv \langle X| \mathcal{O}_L |X\rangle$.

Generically, these classical solutions may spontaneously break (some part of) the global symmetry and are therefore parametrized by a set of moduli $\{c_i\}$ describing the action of (a subgroup of) $G$ on the solutions. Beginning from a given solution $X^*_0$, one can generate a moduli space of solutions under an orbit of the $G$-action $X^*_0 \to X^*_{\{c_i\}}$. Since these solutions contribute equal exponential factors, one must integrate over this moduli space in order to reproduce the correct saddle-point approximation to the correlator. Additionally, in the case where $J$ and $J'$ are not equal, and $J-J' \ll 1$, one needs to take into account the contributions coming from the WKB wavefunction of the initial and final states
\be
    \bra{J'}\ket{X^*_c}\approx e^{-iJ' c}, \qu \bra{X^*_c}\ket{J}\approx e^{iJ c}.
\ee The condition  $J-J' \ll 1$ is necessary for the WKB approximation to hold.
Putting it all together, the semiclassical correlator is given by the orbit average
\be
    \langle J'| \mathcal{O}_L(t=0) | J \rangle\approx e^{iS\[X^*\]}\int \prod_i dc_i \; \mathcal{O}_L[X^*_{\{c_i\}}] \;e^{i\left(J_i-J_i'\right)c_i},
\ee
where $\mathcal{O}_L[X^*]$ should be understood as the classical analog of the operator $\mathcal{O}_L$.

\section{BPS Coherent States}\label{3}
In this section we review the coherent state methods introduced in \cite{Berenstein:2022srd}, and their application to BPS correlators of fully-symmetric Schur polynomials. Firstly, half-BPS operators in $\mathcal{N}=4$ SYM are described by polynomials in the traces of a complex scalar field $Z$. For our purposes we will want to consider the theory on the cylinder $\mathbb{R}\times S^3$, so that our initial and final states are inserted at $t=\pm \infty$.  Then, the main idea is that the following expression serves as a generating series for all half-BPS operators
\begin{equation}\label{HCIZ}
    \ket{\Lambda}= \frac{1}{\texttt{Vol}\left(U(N)\right)}\int_{U(N)}dU e^{\Tr\left( U\Lambda U^\dagger Z\right)}\ket{0},
\end{equation}
where $\Lambda$ is a diagonal matrix with complex eigenvalues $\lambda_i$. A simple calculation shows that this is in fact a coherent state, in the sense that the action of $\bar{Z}$ on this state can be replaced by multiplication by $U\Lambda U^\dagger$. This state also has a simple expression as a sum over the Schur basis
\begin{equation}
    \ket{\Lambda}= \sum_{R}\frac{1}{d_R} \chi_R\left(\Lambda\right)\chi_R(Z)\ket{0},
\end{equation}
where $d_R$ is the norm of the state created by $\chi_R(Z)$. One important property of this formalism is that correlation functions involving these coherent states can be recast in terms of a unitary matrix integral. For example, the overlap between two coherent states has an explicit formula as a sum over saddle points
\begin{equation} \label{saddlesum}
    \bra{\bar{\Lambda}}\ket{\Lambda}= \int_{U(N)}dU e^{\Tr\left( U\Lambda U^\dagger \bar{\Lambda}\right)}=\mathcal{C}_N \sum_{\sigma \in S_N}(-1)^{\text{sign}(\sigma)}\frac{e^{\sum_i\lambda_i \bar{\lambda}_{\sigma(i)}}}{\Delta(\Lambda)\Delta(\bar{\Lambda})};
\end{equation}
 the overall normalization constant $\mathcal{C}_N$ that depends on conventions.  More generally, commuting the exponential in $\ket{\Lambda}$ with insertions of $\bar{Z}$ will have the effect of replacing $\bar{Z}$ with $\bar{Z}+U\Lambda U^\dagger$, and similarly for $Z$.  Although this formulation is quite explicit, it is unclear that the term with the largest exponential actually dominates the sum, since there are $N!-1$ other saddle point contributions that could in principle lead to an exponentially large correction. This is not always the case, since the leading contribution always corresponds to the identity permutation, while the remaining saddle points are weighted by a sign; whenever some of the eigenvalues $\{\lambda_i\}$ are exponentially close to one another different saddles become comparable to the identity saddle and their contribution will become important. 
 
 For our purposes, we will restrict to the case where a single eigenvalue $\lambda_1=\lambda$ is taken to be non-zero, which restricts the sum over representations to those associated with single row Young diagrams. In this case, the formula in terms of unitary integrals is difficult to perform calculations with simply because the numerator and denominator in \eqref{saddlesum} become degenerate. To remedy this, one should realize that the integral is really being performed over an orbit parametrized by:
 \begin{equation}
     \mathcal{O}_{\lambda}=\{\lambda UP_1 U^\dagger\;\;| \;\;U\in U(N)\}\cong \mathbb{CP}^{N-1},
 \end{equation}
 where $P_1$ denotes the projector into the eigenspace of $\lambda_1$. Geometrically this is straightforward to understand; the projector operator $P_1$ is naturally associated to a unit vector in $\mathbb{C}^N$ by 
 \begin{equation}
     P_1= \varphi \varphi^\dagger,
 \end{equation}
 so a choice of projector $P_1$ is in one-to-one correspondence to a line in $\mathbb{C}^N$. The action of the unitary group moves this line inside of $\mathbb{C}^N$, so the resulting integral should be taken over the Grassmannian $\mathbf{Gr}(1,N)\cong\mathbb{CP}^{N-1}$. This suggests that the natural generalization of \eqref{HCIZ} is actually
 \begin{equation}
   \ket{\lambda}=  \frac{1}{\texttt{Vol}(\mathbb{CP}^{N-1})}\int_{\mathbb{CP}^{N-1}} d\varphi^\dagger d\varphi \;e^{\sqrt{N}\lambda\varphi^\dagger Z\varphi}\ket{0}.
 \end{equation}
    Formally this generating functions looks similar to the generating function introduced in \cite{Chen:2019gsb}, but there are many important differences that make $\ket{\lambda}$ much better behaved.
    
    The first important difference is that $\ket{\lambda}$ is a coherent state of finite norm:
    \begin{equation}
    \begin{aligned}
       \Tr\left[\bar{Z}^L\right]\ket{\lambda}&=\lambda^L\ket{\lambda}      \\
    \end{aligned}
    \end{equation}
To find the norm of this state, we can exploit the fact that the integration measure on $\mathbb{CP}^{N-1}$ is left-invariant under the action of $U(N)$. More precisely, once we use the Baker-Hausdorff-Campbell formula to commute the exponentials coming from $\bra{\lambda}$ and $\ket{\lambda}$ we are left with a pair of integrals as in \cite{Jiang:2019xdz,Chen:2019gsb}:
\begin{equation}
   \bra{\lambda}\ket{\lambda}= \left(\frac{1}{\texttt{Vol}(\mathbb{CP}^{N-1})}\right)^2\int_{\left(\mathbb{CP}^{N-1}\right)^2}d\varphi^\dagger d\varphi \;d\psi^\dagger d\psi\;e^{N \bar{\lambda}\lambda\varphi^\dagger \psi \psi^\dagger\varphi}.
\end{equation}
At this point our analysis differs from that of \cite{Jiang:2019xdz,Chen:2019gsb,Yang:2021kot}, in that we can proceed without performing a Hubbard-Stratonovich transformation. To see why this is the case, we should remember that we may parametrize $\psi$ as a rank-one projector conjugated by a unitary matrix $\psi= U P_1 U^\dagger$. Since the measure for $\phi$ is invariant, this reduces the integral over $\psi$ into a volume integral over $U(N)/\(U(N-1)\times U(1)\) \cong \mathbb{CP}^{N-1}$:
\begin{equation}
\begin{aligned}
   \bra{\lambda}\ket{\lambda}&= \frac{1}{\texttt{Vol}\left(\mathbb{CP}^{N-1}\right)}\int_{\mathbb{CP}^{N-1}}d\varphi^\dagger d\varphi \;e^{N \bar{\lambda}\lambda\varphi_1^* \varphi_1}\\
   &= \frac{1}{\texttt{Vol}\left(\mathbb{CP}^{N-1}\right)}\int_0^1 dr (1-r)^{N-1} e^{N \bar{\lambda}\lambda r}\\&= \frac{(N-1)^!}{\left(N\pi\right)^{N-1}}\sum_{L=N-1}^\infty \frac{\left(N\bar{\lambda}\lambda \right)^L}{L!}\simeq \sqrt{N}e^{N\bar{\lambda}\lambda},
   \end{aligned}
\end{equation}
where we have chosen the projective space to have radius $\sqrt{N}$, and used the Stirling approximation in the last line. In comparison, the norm of the state created by a determinant operator has a norm given by \cite{Chen:2019gsb}
\begin{equation}
\begin{aligned}
    \langle \det\left(\bar{Z}- \bar{\lambda} \right)\det\left(Z-\lambda \right) \rangle&= \int_0^\infty dr\; e^{-N r}(\bar{\lambda}\lambda+r)^N\\
    &= \frac{N!}{N^N}\sum_{k=0}^N \frac{(N|\lambda|^{2})^k}{(N-k)!}\simeq \sqrt{N} e^{N |\lambda|^2}.
\end{aligned}
\end{equation}

Hence, the second important difference between the approach using an inverse determinant operator and $\ket{\lambda}$ is that we do not need to introduce an additional set of auxiliary variables to obtain an integral which we can evaluate via the saddle-point approximation, and the resulting saddle-point equations are equivalent in both approaches. 
The most important difference between our approach is the fact that we can easily generalize our construction to write down generating functions of characters associated to Young diagrams with more than one row in a very compact way, with very explicit formulas. For instance, the product of determinant operators has a character expansion coming from one of the Cauchy identities of Schur functions:
\begin{equation}
   \prod_{i=1}^k \det(Z-\lambda_i)= \det(Z\otimes \mathbf{1}_k-\mathbf{1}_N\otimes \Lambda_k)=\det(\Lambda_k)^N\sum_{R} \chi_R(Z) \chi_{R^T}(-\Lambda_k^{-1}),
\end{equation}
and there is a similar expansion for inverse determinants
\begin{equation}
   \prod_{i=1}^k \det(Z-\lambda_i)^{-1}= \det(Z\otimes \mathbf{1}_k-\mathbf{1}_N\otimes \Lambda_k)^{-1}=\det(\Lambda_k)^N\sum_{R} \chi_R(Z) \chi_R(-\Lambda_k^{-1}).
\end{equation}
One difficulty with dealing with the latter expression is that once one performs the Hubbard-Stratonovich transformation the resulting saddle-point equations are complicated matrix equations, and extracting the contribution from each character seems difficult. Also, the state created by such an operator does not have a finite norm, so the expressions have to be treated as formal generating functions. In our approach, one can generate the same class of states by letting $\Lambda$ in \eqref{HCIZ} have $k$ non-zero eigenvalues and integrating over the appropriate homogeous space. More precisely, one needs to replace the integral over $\varphi$ by an integral over an isometry
\begin{equation}
\begin{aligned}
V V^\dagger&= P_k\\
    V^\dagger V&= \mathbf{1}_n,
\end{aligned}
\end{equation}
where $P_k$ is a rank $k$ projector. The integration is then performed over the space of $k$-dimensional subspaces of $\mathbb{C}^N$,  which is the Grassmannian $\mathbf{Gr}(k, N)$. Similar types of integrals have been studied previously in the literature \cite{Fujii:1995ds}, and they are known to have exact formulas in terms of iterated residues \cite{MagdalenaZielenkiewicz2014}. 
\section{Gauge Theory Computation}\label{4}
\subsection{Generating function for BPS Three-Point Functions}
We are interested in computing the overlap of two AdS giant graviton states with a light BPS single trace operator. This correlator is related to the three-point structure constant by
\be \label{eq:3pt_gauge}
C_{\mathcal{S}_{J'} \mathcal{S}_J \mathcal{O}_L}=\frac{\bra{\mathcal{S}_{J'}} \Tr\left[\tilde{Z}^L \right]\ket{\mathcal{S}_J}}{\sqrt{L\bra{\mathcal{S}_{J'}}\ket{\mathcal{S}_{J'}}\bra{\mathcal{S}_{J}}\ket{\mathcal{S}_{J}}}},
\ee
where $\tilde{Z}$ is the twisted translated frame operator $\tilde{Z}=\frac{Z+\Bar{Z}+ (Y-\Bar{Y})}{2}$. The boundary states $\ket{\mathcal{S}_J}$ are to be understood as the insertion of a fully-symmetric Schur polynomial operator\footnote{We differentiate between the notation used in \cite{Jiang:2019xdz, Yang:2021kot}, $\ket{\mathcal{D}_J}$, since the operators we are considering are not determinants.} at $t=\pm \infty$
\begin{equation}
    \ket{\mathcal{S}_J}= \chi_{(J)}(Z)\ket{0}.
\end{equation}
Because half-BPS correlators are protected, we can perform the calculations in the free field theory.
The boundary states can be generated using the following operators:
\begin{equation}
\begin{aligned}
\ket{\lambda}&= \int_{\mathbb{CP}^{N-1}} d\varphi e^{\sqrt{N-1}\lambda \varphi^\dagger Z \varphi}\ket{0}\\
\bra{\Lambda}&= \bra{0}\int_{U(N)} dU e^{\sqrt{N-1}\Tr\left(\Bar{Z}U^\dagger \Bar{\Lambda} U\right)},
\end{aligned}
\end{equation}
where we can set $\Bar{\Lambda}= \Bar{\lambda} P_1$ during the later parts of the computation. The advantage of this setup is that the measure $d\phi$ is invariant under unitary transformations, so when we commute the exponentials using the Campbell-Hausdorff formula the integral over the unitary group will drop out of the correlator. The overlap that we will want to compute is given by:
\begin{equation}
      \mathcal{F}(\lambda, \Lambda, t)=  \bra{\Lambda} \Tr\left[\frac{1}{1-2t \tilde{Z}}\right]\ket{\lambda}.
\end{equation}

When we commute all raising and lowering operators past each other, the net effect is to replace the the fields by their saddle-point value
\begin{equation}
\mathcal{Z}\rightarrow \frac{U^\dagger \Bar{\Lambda}U+ \lambda\varphi\varphi^\dagger }{2}= U^\dagger\left(\frac{ \Bar{\Lambda}+\lambda U\varphi\varphi^\dagger U^\dagger }{2}\right)U.
\end{equation}
Since the expression inside the exponential only depends on $U\varphi$ after applying the Campbell-Hausdroff formula, all unitaries can be reabsorbed by a change of variables. So in the end the generating function $\mathcal{F}$ is expressed as:

\begin{equation}
       \mathcal{F}(\lambda, \Lambda, t)= \int_{\mathbb{CP}^{N-1}} d\varphi \, e^{(N-1)\lambda \varphi^\dagger \Bar{\Lambda} \varphi}\Tr\left[\frac{1}{\textbf{1}-t\left(\Bar{\Lambda}+\lambda \varphi\varphi^\dagger  \right)}\right].
\end{equation}
This integral can be computed exactly via equivariant localization. A simple way of seeing this is that the integral may be turned into a Gaussian integral subject to the constraint $|\varphi|^2=1$.
\begin{equation}
    \delta(|\varphi|^2-1)= \int ds \; e^{\lambda s(|\varphi|^2-1)}.
\end{equation}
After this substitution, we can perform the Gaussian integral over $\varphi$ on the whole complex plane by contour integration. By choosing a set of contours such that the phase of the exponential is stationary, the resulting integrals are Gaussian integrals peaked at the eigenvalues of $\Bar{\Lambda}$, so in the end we only need to sum over $N$ saddle points. After performing the Gaussian integral, each saddle point will correspond to a pole on the complex $s$ plane, and every insertion of $\varphi_i \varphi^\dagger_j$ in the integral can be replaced by its moment taken from the Gaussian distributions;
\begin{equation}
    :\varphi_i \varphi^\dagger_j: \;\sim \left(\frac{1}{\lambda(s-\Bar{\Lambda})}\right)_{ij}.
\end{equation}
After this we are left to compute a contour integral over the complex $S$ plane over a infinitely large circle.  The particular choice of orientation for the contours that we need guarantees that the sum in homology of the $N$ contours is equivalent to the trivial contour encircling a pole at infinity. 

In practice we will need consider cases where $\bar{\Lambda}$ has rank one, which makes the torus action on $\mathbb{CP}^{N-1}$ degenerate. However, we can still compute the integral exactly as a sum over residues of poles of higher order. When $\bar{\Lambda}$ has one non-zero eigenvalue, the integral is exponentially dominated by a single saddle point. 

In order to compute the integral with the resolvent, we will need to invert the matrix inside of the trace. A simple way of doing this is by writing this matrix as 
\begin{equation}
  \textbf{1}-t\left(\Bar{\Lambda}+\lambda \varphi\varphi^\dagger  \right)=  \textbf{1}- \Phi_i \Sigma_{ij}\Phi^\dagger_j,
\end{equation}
where $\Sigma$ is a $2\times 2$ diagonal matrix with components $(t\bar{\lambda}, t\lambda)$, and $\Phi_i$ is an $N\times2$ matrix consisting of $(v_1, \varphi)$, where $v_1$ is a unit vector. The inverse of this matrix is given by
\begin{equation}
   \left( 1-t\left(\Bar{\Lambda}+\lambda \varphi\varphi^\dagger  \right)\right)^{-1}= 1+\Phi \left( \Sigma^{-1}+ \Phi^\dagger \Phi \right)^{-1}\Phi^\dagger.
\end{equation}
In some respects, the matrix $\Sigma_{ij}$ plays a similar role as the Hubbard-Stratonovich field $\rho$ needed to simplify correlation functions involving determinants. 
When we take the trace the first term will be independent of $t$, so it will not contribute to the three-point and the second term becomes a trace over the $2\times 2$ auxiliary indices
\begin{equation}
\begin{aligned}
    \Tr \left( 1-t\left(\Bar{\Lambda}+\lambda \varphi\varphi^\dagger  \right)\right)^{-1}&= N+\tr\left(\Phi^\dagger\Phi \left( \Sigma^{-1}+ \Phi^\dagger \Phi \right)^{-1}\right)\\  
   &= N-2\left(\frac{1-\frac{t}{2}\left(\lambda+ \bar{\lambda} \right)}{t^2 \lambda \bar{\lambda}(\varphi_1^*\varphi_1-1)+t\left(\lambda+ \bar{\lambda} \right)-1}+1\right).
\end{aligned}
\end{equation}
So the exact expression for the form factor is obtained by performing the integral over $\varphi$ and since the first and third terms are analytic in $t$ we can simply ignore them:
\begin{equation}
   \mathcal{F}(\lambda, \bar{\lambda}, t)\simeq -\int_{\mathbb{CP}^{N-1}} d\varphi \, e^{(N-1)\lambda\Bar{\lambda} \varphi_1^*  \varphi_1}\left(\frac{1-\frac{t}{2}\left(\lambda+ \bar{\lambda} \right)}{t^2 \lambda \bar{\lambda}(\varphi_1^*\varphi_1-1)+t\left(\lambda+ \bar{\lambda} \right)-1}\right).
\end{equation}
This integral can be evaluated easily by using spherical coordinates and expanding in powers of $\varphi_1^*\varphi_1$, but it will turn out to be better to approximate this quantity via the saddle-point approximation.
\subsection{Large $N$ Limit}
    The first thing to note about the integral expression for $\mathcal{F}$ is that the integrand breaks the $U(N)$ symmetry of the measure to $U(1)\times U(N-1)$, so it is convenient to perform the angular integration over the $N-1$ directions perpendicular to $\varphi_1$ first. For a fixed value of $\varphi_1$, this is given by half of the volume of a sphere of radius $(1-\varphi_1^*\varphi_1)^{1/2}$. The exact value of the angular integrals is not very important since the overall factor in front of the generating function will cancel when we normalize the structure constants, but what is important is that the integral over $\varphi_1$ is done with the correct measure
\begin{equation}
    \mathcal{F}(\lambda, \Lambda, t)\simeq- \frac{\pi^{N-2}}{(N-2)!}\int drd\vartheta (1-r)^{N-1}e^{(N-1)\lambda\Bar{\lambda} r}\left(\frac{1-\frac{t}{2}\left(\lambda+ \bar{\lambda} \right)}{t^2 \lambda \bar{\lambda}(r-1)+t\left(\lambda+ \bar{\lambda} \right)-1}\right).
\end{equation}
Finally, we can evaluate this integral using the saddle-point approximation. Since the terms coming from the resolvent do not scale with $N$, they will not lead to large exponents, so only need to consider the critical points of the following effective action
\begin{equation}
    S_{\texttt{eff}}= \lambda \bar{\lambda}r + \log\left( 1-r\right).
\end{equation}
The saddle points of this action precisely fix $r$ in such a way as to simplify the denominator of the expression in parentheses:
\begin{equation}
    \lambda \bar{\lambda}= \frac{1}{(1-r)},
\end{equation}
which yields
\begin{equation}
    \mathcal{F}_{\texttt{saddle}}(\lambda, \bar{\lambda}, t) = \left(\frac{1-\frac{t}{2}\left(\lambda+ \bar{\lambda} \right)}{t^2 -t\left(\lambda+ \bar{\lambda} \right)+1}\right) e^{S_{\texttt{saddle}}}.
\end{equation}
Since the exponential factor computes the saddle point value of the overlap of the AdS giant states it will cancel when we compute the structure constants so we will omit its explicit form.
\subsection{Diagonal Structure Constants}
We now have an approximate expression for the form factor $\mathcal{F}(\lambda, \bar{\lambda}, t)$ which is a generating function for three point functions involving two BPS AdS giant gravitons and  a BPS singe trace operator. One feature of this calculation is that the form factor describes a semiclassical giant graviton localized along some null geodesic on the hemisphere of $S^5$. To obtain a correlator with fixed $R$ charge we need to average over the position of the giant to project into a fixed charge state. In our case, the moduli of the solution is the phase of the eigenvalue $\lambda$. So it is natural that we perform an average over the orbit generated by the phase of $\lambda$
\begin{equation}\label{generating function}
\begin{aligned}
      \lambda&= y\cosh \rho_0 \\
      \bar{\lambda}&=\frac{1}{y} \cosh \rho_0 ,
\end{aligned}
\end{equation}
which gives
\begin{equation}
 \mathcal{G}(t)= -\frac{1}{2\pi i}\oint\frac{dy}{y} \left[\frac{1-\frac{t}{2}\left(y+ \frac{1}{y} \right)\cosh \rho_0}{t^2 -t\left(y+ \frac{1}{y} \right)\cosh \rho_0-1}\right]= -\frac{1-t^2}{ \sqrt{t^4-2
   t^2 \cosh 2 \rho +1}}.
\end{equation}
To obtain the one-point function of a BPS single trace operator we simply expand this function in $t$, and extract the $L$'th coefficient with a contour integral:
\begin{equation} \label{eq:diag_gauge}
\begin{aligned}
     C_{\mathcal{S}_\Delta \mathcal{S}_{\Delta}\mathcal{O}_L}&=  \frac{1}{2\pi i \sqrt{L}} \oint\frac{dt}{t^{L+1}}\mathcal{G}(t)= -\sum_{J=0}^\infty \frac{1}{2\pi i\sqrt{L}} \oint\frac{dt}{t^{L+1}} P_{J}\left(\cosh 2 \rho_0\right) t^{2J}(1-t^2) \\
    &=-\frac{1^L+(-1)^L}{2\sqrt{L}}\left( P_{\frac{L}{2}}\left(\cosh 2 \rho_0\right)-P_{\frac{L}{2}-1}\left(\cosh 2 \rho_0\right)\right).
\end{aligned}
\end{equation}
This is exactly the answer obtained in \cite{Yang:2021kot}, with a minor difference. In their analysis one needs to perform an integral over $|\lambda|$ with a measure that effectively replaces it with the discrete dimension of the operator dual to the AdS giant. In our case we are still left with $\cosh 2\rho_0$ as a continuous parameter corresponding to the radial position of the brane inside AdS. This will match exactly the answer obtained from the semiclassical computation. 

\subsection{Off-Diagonal Structure Constants}
Since our integral formula is identical to the one found in \cite{Yang:2021kot}, we can borrow their results to find the off-diagonal structure constants. The idea is to replace the integral over the phase in \eqref{generating function} by 
\begin{equation}
    \oint \frac{dy}{iy}\longrightarrow  \oint \frac{dy}{iy^{k+1}}.
\end{equation}
This is the contribution from the wavefunctions of the boundary states whenever the difference of the $R$-charges of the in and out states is $k$. If $k\ll N$, the saddle point is not modified, and the integrand remains the same. The structure constant for AdS giant gravitons can also be obtained by analytically continuing the structure constant for sphere giant gravitons
\begin{equation}
    \theta_0\rightarrow i\rho_0+ \pi/2.
\end{equation}
Here we will prove the formula given in \cite{Yang:2021kot} for all values of $k$. After performing the residue integral over $y$ we arrive at the following expression for the generating function of off-diagonal structure constants at large $N$
\begin{equation}
\begin{aligned}
\mathcal{G}_k(t)&=-\frac{1}{2} \frac{\left(t^k (t^2-1) \cosh^k\rho_0 \right)2^{k}}{\sqrt{t^4-2
   t^2 \cosh 2 \rho +1}\left(1+t^2+\sqrt{t^4-2
   t^2 \cosh 2 \rho +1} \right)^k}\\
  &=\frac{1}{2} t^k (1-t^2)\sum_{J=0}^{\infty} P_{J}^{(0,k)}\left(\cosh 2 \rho_0 \right) t^{2J},
\end{aligned}
\end{equation}
where we used the generating function for Jacobi polynomials\footnote{The Jacobi polynomials $P^{(\a,\b)}_n(x)$ span a large family of orthogonal polynomials; they reduce to Legendre polynomials for $\a = \b = 0$. For  $\a = 0$ and $\b=k$ they correspond to the radial parts of Zernike polynomials.} $P_{J}^{(\alpha,\beta)}(x)$ to expand the function in powers of $t^2$. Finally, we perform the contour integral over $t$ to obtain the off-diagonal structure constant.
\begin{equation}\label{eq:offdiag_gauge}
\begin{aligned}
     C_{\mathcal{S}_{\Delta+k} \mathcal{S}_{\Delta}\mathcal{O}_L}&=  \frac{1}{2\pi i \sqrt{L}} \oint\frac{dt}{t^{L+1}}\mathcal{G}_k(t)\\
    &=-\frac{1^{L-k}+(-1)^{L-k}}{2\sqrt{L}}\times \cosh^k\rho_0\left( P_{\frac{L-k}{2}}^{(0,k)}\left(\cosh 2 \rho_0\right)-P_{\frac{L-k}{2}-1}^{(0,k)}\left(\cosh 2 \rho_0\right)\right).
\end{aligned}
\end{equation}
Similarly, the formula for the off-diagonal structure constants for sub-determinant operators can be written as
\begin{equation}
  C_{\mathcal{D}_{\Delta+k} \mathcal{D}_{\Delta}\mathcal{O}_L}
    =-\frac{i^{L-k}+(-i)^{L-k}}{2\sqrt{L}}\times \sin^k\theta_0\left( P_{\frac{L-k}{2}}^{(0,k)}\left(\cos 2 \theta_0\right)+P_{\frac{L-k}{2}-1}^{(0,k)}\left(\cos 2 \theta_0\right)\right).
\end{equation}

We have checked that our formula agrees with the formula given in \cite{Yang:2021kot} for many values of $L$ and $k$, and the two formulas can be turned into one another by using the recurrence relations of the hypergeometric function. In the extremal limit $L=k$, the second term in both structure constants vanish and the Jacobi polynomials reduce to a factor of unity, so the formula is well-defined for all $k\leq L$. For $k > L$, the contour integral has no poles, and so the structure constants vanish identically as expected from $R$-charge conservation.

\section{Holographic Computation}\label{5}

We now move on to the holographic computation of the structure constant \eqref{eq:3pt_gauge}. To do so, we simply replace each part of the formula by its holographic counterpart. The dual of the Schur polynomial operators $|\mathcal{S}_J\rangle$ are BPS AdS giant gravitons with angular momentum $\Delta=J$, whose quantum state we denote by $|\hat{\mathcal{S}}_\Delta\rangle$. The single trace operator $\Tr\[\wtd{Z}^L\]$ is replaced by an operator $\hat{\mathcal{O}}_L$ which describes the backreaction on the worldvolume of the giant graviton. Altogether the holographic structure constant is given by
\be
C_{\hat{\mathcal{S}}_{\Delta'} \hat{\mathcal{S}}_\Delta \hat{\mathcal{O}}_L}=\frac{\bra{\hat{\mathcal{S}}_{\Delta'}} \hat{\mathcal{O}}_L \ket{\hat{\mathcal{S}}_\Delta}}{\sqrt{L\bra{\hat{\mathcal{S}}_{\Delta'}}\ket{\hat{\mathcal{S}}_{\Delta'}}\bra{\hat{\mathcal{S}}_{\Delta}}\ket{\hat{\mathcal{S}}_{\Delta}}}}.
\ee
The three-point function can be computed from a path integral on the worldvolume of the giant graviton, which is amenable to a saddle-point analysis. As we will see, a proper treatment via the orbit average method will yield a result which matches the gauge theory exactly.

\subsection{AdS Giant Graviton Solution}
We will be interested in solutions to the DBI action describing a giant graviton wrapping an $S^3\subset$ AdS$_5$, which rotates along the equator of the $S^5$ at the speed of light. For our set up it will be convenient use global coordinates to parametrize AdS$_5\times S^5$:
\be
    ds^2= -\cosh^2\rho dt^2+ d\rho^2+\sinh^2\rho d \wtd{\Omega}_3^2 + d\Omega_5^2,
\ee
where the metric of the five-sphere is 
\be
    d \Omega_5^2= d\theta^2 + \sin^2 \theta d\phi^2+ \cos^2\theta \left( d \chi_1^2+\sin^2\chi_1d\chi_2^2+\cos^2 \chi_1 d\chi_3^2\right).
\ee
We can then gauge fix the worldvolume coordinates $\sigma^\mu$ of the D-brane to agree with the coordinates of $\mathbb{R}_t\times S^3 \subset \text{AdS}_5$
\begin{equation}
    \r = \r_0,\;\; \; \s^0 = t = \p, \;\;\; \s^i = \wtd\c_i
\end{equation}
where the tilded coordinates $\wtd\chi_i$ are the coordinates of the three-sphere inside AdS$_5$. The size of a BPS giant graviton is equal to its R-charge (angular momentum along $S^5$), which is related to its radial position in the AdS direction by
\begin{equation}
    \cosh\rho_0= \frac{J}{N}, \;\;\;J\geq N.
\end{equation}
To compute the three-point function we will need to compute the corrections to the D3-brane action coming from a light supergravity perturbation as in \cite{Berenstein:1998ij, Lee:1998bxa}.

\subsection{Fluctuations of the D3-brane action}
The action for an AdS giant graviton is given by the sum of the DBI and Wess-Zumino (WZ) actions
\begin{equation}
    S = -\frac{N}{2 \pi^2}\int d^4 \sigma \left( \sqrt{-h}+ P[C_4] \right),
\end{equation}
where $h$ is the induced worldvolume metric and $P[C_4]$ is the pull-back of the Ramond-Ramond four-form potential of the background. For our purposes we will want to concentrate on the RR flux through the AdS factor
\be
    C_4= -\sinh^4\rho dt\wedge\text{Vol}(\wtd{\Omega}_3).
\ee
The light operator insertion can then be identified by the perturbations to the D3-brane action \cite{Costa:2010rz}
\be
    \hat{\mathcal{O}}_L = \delta S_\text{DBI} + \delta S_\text{WZ}.
\ee
For this we will need the fluctuations of the spacetime metric $g$ as well as for the four-form potential \cite{Kim:1985ez}:
\bea
&&\delta g_{\mu \nu}= \left[-\frac{6}{5}L g_{\mu \nu}+ \frac{4}{L+1} \nabla_{(\mu} \nabla_{\nu)}\right] s^L(X)Y_L(\Omega_5)\\
&&\delta g_{\alpha \beta}= 2 L g_{\alpha \beta} s^L(X) Y_L(\Omega_5)\\
&&\delta C_{\mu_1 \mu_2 \mu_3 \mu_4}= -4 \epsilon_{\mu_1 \mu_2 \mu_3 \mu_4 \mu_5} \nabla^{\mu_5} s^L(X) Y_L(\Omega_5),
\eea
where $\m, \n, \dots $ denote coordinates on the AdS$_5$, $\a, \b, \dots $ denote coordinates on the $S^5$, $Y_L(\Omega_5)$ denotes a spherical harmonic on the $S^5$, and $s^L(X)$ is the bulk-to-boundary propagator. The kinds of fluctuations that are dual to the operator $\Tr\left[\wtd{Z}^L\right]$ are given by choosing a spherical harmonic corresponding to the homogeneous polynomial $(Z+\bar{Z}+ Y- \bar{Y})^L$, where $X$, $Y$, $Z$ are the coordinates on $S^5\subset\mathbb{C}^3$,
\begin{equation}
    Y_L(\wtd{Z})= \left(\sin \theta \cos \phi +i \cos \theta \cos \chi_1 \sin \chi_3 \right)^L.
\end{equation}
The bulk-to-boundary propagator is given by
\begin{equation}
    s^L(X)= \frac{\mathcal{N}}{(-2 P\cdot X)^L},
\end{equation}
where $P$ represents the coordinates of the operator insertion on the boundary coordinates $P^I$ and $X$ are the embedding coordinates of AdS$_5$:
\begin{equation}
\begin{aligned}
      X^{-1}&=\cosh \rho \cosh t_E , \;\; X^0= \cosh \rho \sinh t_E\;\;\;X^i = \sinh\rho\;n^i\\
      P^{-1}&= \cosh \bar{t}_E,\;\;\; P^0= \sinh \bar{t}_E, \;\;\; P^i= \bar{n}^i,
\end{aligned}
\end{equation}
where $t_E$ is the Euclidean time coordinate $t_E=i t$ and $|n|^2=|\bar{n}|^2=1$. The unit vectors $n^i$ and $\bar{n}^i$ represent the position of the operator insertion on the $S^3$ inside AdS$_5$ in the bulk and the boundary, respectively. In our case the bulk-to-boundary propagator is given by
\begin{equation}
    s^L(t_E, n^i, \bar{n}^i)= \frac{\mathcal{N}}{2^L}\frac{1}{\left(\cosh \rho_0 \cosh t_E - n \cdot \bar{n} \sinh \rho_0\right)^L},
\end{equation}
where $n \cdot \bar{n} = \cos\wtd\c_1 \sin\wtd\c_3$ and the normalization $\mathcal{N}$ is chosen such that the two-point function is unit-normalized.

\paragraph{DBI Action Fluctuations}
The fluctuation of the induced metric on the D3-brane has the form
\bea
\delta\sqrt{h} = \frac{1}{2}\sqrt{h}h^{ab}\left(\partial_a X^\mu \partial_b X^\nu \delta g_{\mu \nu}+ \partial_a X^\alpha \partial_b X^\beta g_{\alpha  \beta} \right).
\eea
Substituting the worldvolume coordinates into the variation of the induced metric gives 
\begin{equation}\label{fluctuations}
    \delta \sqrt{h}=\frac{1}{2}\sqrt{h}\left( \frac{4}{L+1}h^{ab}\;\nabla_a \nabla_b-\frac{2L(L-1)}{L+1} h^{ab}g_{ab}+2L h^{tt}\right) s^L Y_L.
\end{equation}
To simply this expression, it is useful to exploit the fact that $s^L(X)$ is a scalar field of of mass-squared $L(L-4)$ in AdS units. To use this fact we may rewrite the induced metric on the brane in terms of the metric of AdS$_5$
\begin{equation}
\begin{aligned}
    h_{ab}&= g_{ab}+\delta_a^t\delta_b^t \\
    h^{ab}&= g^{ab}-\(\sinh^2\r \cosh^2\r\)^{-1} \, \delta^a_t \delta^b_t,
\end{aligned}
\end{equation}
and then we can add and subtract the second covariant derivative in the $\rho$ direction to complete the Laplacian in \eqref{fluctuations} which gives:
\ba
\d S_{\text{DBI}} &= \frac{N}{2\pi^2} \int d^3\s \, \d \sqrt{h}|_{t = 0} \\
&= \fr{N}{4\pi^2} \sinh^2\r_0 \int d^3\s \, F_\text{DBI} |_{t = 0}
\ea
where
\bea
F_\text{DBI} &=& -\fr{1}{N}\fr{4}{L+1}\sin\c_1\cos\c_1 \(\fr{\pa_t^2}{\cosh^2\r_0} + \sinh^2\r_0 \pa_{\r}^2 - \tanh\r_0 \pa_\r + L^2 
\cosh^2\r_0 + 2L\sinh^2\r_0 \) s^L Y_L \nn
\eea
The differential operator in parenthesis basically raises the spin of the propagator $S^L$ by two units $L+2$ and multiplies it by a simple polynomial in $\sinh\rho_0$ and $n\cdot \bar{n}$.

\paragraph{WZ Action Fluctuations}
The fluctuations of the WZ term are straightforward to compute. The four-form potential only has indices in the AdS$_5$ directions, so the only possible term is
\bea
\delta C_{t \wtd \c_1 \wtd \c_2 \wtd \c_3} &=& -4\pa_{\r} s^L(X) Y_L(\Omega_5). \nn
\eea
The contribution from the WZ action is thus
\ba
\d S_{\text{WZ}} &= \fr{N}{2\pi^2} \int d^3\s P\[\d C_4\] = \fr{N}{2\pi^2} \int d^3\s \sqrt{g_{\text{AdS}_5}} \delta C_{t \wtd \c_1 \wtd \c_2 \wtd \c_3} \\
&= -\fr{N}{4\pi^2} \sinh^2\r_0  \int d^3\s \, F_\text{WZ}|_{t = 0}
\ea
where
\bea
F_\text{WZ} = \fr{8}{N} \sin\wtd\c_1 \cos\wtd\c_1 \sinh\r_0 \cosh\r_0 \, \pa_{\r} s^L Y_L. \nn
\eea

\paragraph{Operator insertion}
Putting everything together, we obtain an expression for the insertion of the light operator $\hat{\mathcal{O}}_L$ in the semiclassical limit. In practice it is useful to rewrite the resulting expression in terms of $s^{L+2}$ so that the DBI and WZ terms combine nicely. As in \cite{Bissi:2011dc}, the combination of DBI and WZ terms simplifies significantly:
\bea \label{eq:op_insert}
\hat{\mathcal{O}}_L\[X^*_0\] = \d S_{\text{DBI}} + \d S_{\text{WZ}} = \fr{N}{4\pi^2} \sinh^2\r_0 \int d^3\s \, \(F_\text{DBI} - F_\text{WZ}\)|_{t = 0}
\eea
where
\bea
F_\text{DBI} - F_\text{WZ} = -\frac{\sqrt{L}(L+1)}{N} \fr{\sin\wtd\c_1 \cos\wtd\c_1 \cos^L\phi}{\(\cosh\r_0 \cos t - \cos\wtd\c_1\sin\wtd\c_3 \sinh\r_0\)^{L+2}}. \nn
\eea
So our analysis closely mirrors that of \cite{Bissi:2011dc}, with some minor differences in the simplification of the fluctuation analysis.

At this point our analysis will differ importantly; in their set-up, they proceed by substituting the worldvolume solution $\phi=t$ and integrating over the insertion time $t$. Our calculation instead follows the prescription used by \cite{Yang:2021kot}, which means that the coordinates appearing in $\hat{\mathcal{O}}_L\[X^*_0\]$ are not the worldvolume coordinates of the giant graviton, and instead they should be thought of as the coordinates of the insertion of the operator on the sphere wrapped by the giant. This means that we should not set $\phi=t$, but instead we should treat $\phi = \phi_0$ as a moduli of the solution. The second moduli of the solution is associated to the action of the dilatation operator $t\rightarrow t+i\tau_0$, which is different from the Lorentzian time evolution of the fluctuation. 

Concretely, one should replace the unshifted solution $X^*_0$ by the shifted solution $X^*_{\p_0,\t_0}$, which can be obtained by $\p\to\p+\p_0$ and $t\to t+i\t_0$ in \eqref{eq:op_insert}
\bea
\hat{\mathcal{O}}_L\[X^*_{\p_0,\t_0}\] = \fr{N}{4\pi^2} \sinh^2\r_0 \int d^3\s \, \[F_\text{DBI}\(\p_0,\t_0\) - F_\text{WZ}\(\p_0,\t_0\)\]|_{t = 0}
\eea
where
\bea
F_\text{DBI}\(\p_0,\t_0\) - F_\text{WZ}\(\p_0,\t_0\) = -\frac{\sqrt{L}(L+1)}{N} \fr{\sin\wtd\c_1 \cos\wtd\c_1 \cos^L\p_0}{\(\cosh\r_0 \cosh\t_0 - \cos\wtd\c_1\sin\wtd\c_3 \sinh\r_0\)^{L+2}}. \nn
\eea
Since these solutions spontaneously break the rotation and dilatation symmetry of the background, the orbit average method tells us to integrate over the moduli space. As we will see this is the correct prescription for computing the three-point function, and this will also fix the apparent discrepancy found in \cite{Bissi:2011dc}. It will also allow us to compute the off-diagonal three-point functions by including contributions from the boundary wavefunctions, which was inaccessible from their analysis. 

\subsection{Diagonal Structure Constants}
We can now obtain the diagonal structure constant by performing the orbit average of \eqref{eq:op_insert}
\be
     C_{\hat{\mathcal{S}}_{\Delta} \hat{\mathcal{S}}_{\Delta}\hat{\mathcal{O}}_L}= \int_{-\infty}^{\infty} d\tau_0 \int_{0}^{2\pi} \frac{d\phi_0}{2\pi}\; \hat{\mathcal{O}}_L\[X^*_{\p_0,\t_0}\].
\ee
The details are presented in appendix \ref{app:diag_orbit} and the final answer can be written in terms of a hypergeometric function
\begin{equation}
    C_{\hat{\mathcal{S}}_{\Delta} \hat{\mathcal{S}}_{\Delta}\hat{\mathcal{O}}_L}= -\frac{1}{2} \(1^L+\left(-1 \right)^L\) \sqrt{L} \times \frac{\tanh^2 \rho_0}{\cosh^L\rho_0}\;{}_2F_1\(1 + \fr{L}{2}, 1 + \fr{L}{2}, 2, \tanh^2\r_0\).
\end{equation}
This answer is of the same form as the result found in \cite{Yang:2021kot} for sphere giant gravitons. To see the matching with the gauge theory computation one needs to apply the recurrence formulas of the hypergeometric function to write the expression above as a sum of two Legendre polynomials
\begin{equation}
      C_{\hat{\mathcal{S}}_{\Delta} \hat{\mathcal{S}}_{\Delta}\hat{\mathcal{O}}_L}= -\frac{1^L+(-1)^L}{2\sqrt{L}}\left( P_{\frac{L}{2}}\left(\cosh 2 \rho_0\right)-P_{\frac{L}{2}-1}\left(\cosh 2 \rho_0\right)\right).
\end{equation}
Clearly this matches exactly with the gauge theory computaion, and is also a simple analytic continuation of the stucture constant involving two sub-determinant operators and a light single trace.

\subsection{Off-Diagonal Structure Constants}
For the off-diagonal structure constants with $\Delta+k\sim \Delta$, the only change to the computation is the contribution from the phases of the boundary wavefunctions:
\bea
C_{\hat{\mathcal{S}}_{\Delta+k} \hat{\mathcal{S}}_{\Delta}\hat{\mathcal{O}}_L}&=& \int_{0}^{2\pi} \frac{d\phi_0}{2\pi} \int_{-\infty}^{\infty} d\tau_0\; \hat{\mathcal{O}}_L\[X^*_{\p_0,\t_0}\]e^{ik\p_0} e^{-k\t_0}.
\eea
The final expression is a simple generalization of the diagonal case, and involves a similar type of hypergeometric function:
\bea
C_{\hat{\mathcal{S}}_{\Delta+k} \hat{\mathcal{S}}_{\Delta}\hat{\mathcal{O}}_L} &=& -\fr{1}{2}\(1^{L-k}+\(-1\)^{L-k}\) \sqrt{L} \\
&& \hspace{-15mm} \times  \fr{\tanh^2\rho_0}{\(\cosh\rho_0\)^{L}} \, {}_2F_1\(1 + \frac{L-k}{2}, 1 + \frac{L+k}{2}, 2, \tanh^2 \r_0\) \nn
\eea
for $k \leq L$, and zero for $k > L$. An analogous computation as in the diagonal case shows that this is equivalent to
\bea
C_{\hat{\mathcal{S}}_{\Delta+k} \hat{\mathcal{S}}_{\Delta}\hat{\mathcal{O}}_L} &=& -\frac{1^{L-k}+(-1)^{L-k}}{2\sqrt{L}}\times \cosh^k\rho_0\left( P_{\frac{L-k}{2}}^{(0,k)}\left(\cosh 2 \rho_0\right)-P_{\frac{L-k}{2}-1}^{(0,k)}\left(\cosh 2 \rho_0\right)\right) \hspace{15mm}
\eea
which matches the gauge theory computation. The details are in appendix \ref{app:offdiag_orbit}.

\subsection{No ambiguities for AdS Giants}
We note that unlike the case of the sphere giant graviton in \cite{Yang:2021kot}, our expression is unambiguous for the extremal case $L = k$. Here we will argue that the extremal case for sphere giant gravitons is also unambiguous if one performs the integrals in the correct order.

Naively, if one computes the extremal case as a limit $k\rightarrow L$ one obtains a spurious divergence coming from an integral over $\phi_0$, yet when one evaluates the integral explicitly the answer is manifestly finite. The seemingly problematic integral in our case is 
\begin{equation}
  \int_{0}^{2\pi}  \frac{d \phi_0}{2 \pi} e^{i L\phi_0} \cos^L\phi_0=\frac{1}{2^L}.
\end{equation}
Clearly the integral is finite and well-defined. However, this integral has an alternate expression in terms of sums of hypergeometric functions with a spurious singularity at $k=L$. If the integral is split into this form, the answer appears to be ambiguous in the sense that it is a sum of two infinite quantities, even though the integral has a well-behaved limit. In contrast, the off-diagonal structure constants for sphere giant gravitons appear to have a real divergence in the extremal limit coming from the average over $\tau_0$, which is multiplied by a prefactor that vanishes in the extremal limit.  In \cite{Yang:2021kot} it was argued that the analytic continuation of the non-extremal case to $k=L$ is ambiguous due to the fact that one can always multiply the result by an analytic function that only modifies the function the behavior of the three-point function at $k=L$. Since there no clear constraints on the analytic properties of the three-point function as a function of $k$, there is no unique analytic continuation of the three-point function. However, in their analysis they separated the expressions in the integrals into a finite piece and an infinite piece multiplied by a zero prefactor. Strictly speaking this is not correct, since the integral is not convergent and depending on how one separates the terms one can obtain different answers for the regularized integral.

Upon closer inspection, the source of the divergence can be traced back to the imaginary part of a factor in the sum of the DBI and WZ terms
\be
 F_{\text{WZ}}-F_{\text{DBI}} = \frac{\sqrt{L}(L+1)}{2N} Y^{L-1}\times\frac{ \left(i \cos \theta \cos \chi_1 \sin \chi_3 \cosh2\tau_0-\cos\phi_0 \sin\theta_0 \right)}{\cosh^{L+2}\tau_0}.
\ee
Note that the first term in the parentheses will also lead to a divergent quantity when we average over $\tau_0$, but this way of splitting the integrals into divergent and finite parts is different from the one used in \cite{Yang:2021kot}. Our expression comes from adding and simplifying both contributions to the fluctuations of the action. Now, if we evaluate the average of the second (finite) term we obtain
\begin{equation}
  C^{\;\texttt{finite}}_{\hat{\mathcal{D}}_{\Delta+L} \hat{\mathcal{D}}_{\Delta}\hat{\mathcal{O}}_L}= -\frac{\sin^L\theta_0}{\sqrt{L}},
\end{equation}
which is exactly the extremal structure constant evaluated in the Schur basis. The remaining term is ambiguous, since its contribution is regularization dependent. A natural choice of regularization is to perform the integral over $\phi_0$ before the $\tau_0$ integral, or equivalently to perform the $\tau_0$ integral with a finite upper and lower bound $\pm T$ and then take the limit $T\rightarrow\infty$.  With this choice the problematic term vanishes and the holographic computation agrees with the field theory computation. Physically this makes sense, since the integral over $\phi_0$ of the first term vanishes due to $R$-charge conservation. Hence if one treats the integrals carefully, there is no ambiguity in defining the three-point functions for sphere giants. In fact similar ambiguities happen for the case where $k>L$ in both computations; if one computes the $\tau_0$ integral first the answer has divergent terms, even though the integral vanishes since the integral over $\phi_0$ is zero.

\section{Discussion}
We computed diagonal and off-diagonal structure constants of two AdS giant gravitons and a light supergravity mode in the large $N$ limit, both in (free) $\mathcal{N}=4$ SYM theory and holographically in $\text{AdS}_5\times S^5$. Our analysis shows a precise matching between both descriptions as expected, even in the cases where ambiguities were believed to appear. A crucial step in our calculations was the orbit average over the moduli space of solutions which spontaneously break the rotation and dilatation symmetry of the AdS$_5\times S^5$ background, and the order in which these integrals are performed is crucial for agreement in the computations of extremal correlators. It would be interesting to apply these methods to the class of open strings solutions \cite{Bak:2011yy} found in \cite{Berenstein:2020jen}, where there appear to be discrepancies between the boundary conditions for the semiclassical string and the spin chain descriptions \cite{deMelloKoch:2007nbd}. Since the positions of the open string endpoints along the worldvolume of the giant appear as extra moduli, one should in principle integrate over them in order to compare with the gauge theory computations. This would explain why certain angular momentum modes are allowed on the spin chain description, even though semiclassically they are forbidden by the boundary conditions.

Our calculation in the gauge theory demonstrate the power of the methods introduced in \cite{Berenstein:2022srd, Holguin:2022drf, Lin:2022wdr} for computing correlators of fully-symmetric Schur polynomials. Our methods are in many respects more streamlined when compared to the approach introduced in \cite{Chen:2019gsb} for dealing with symmetric Schur functions. In principle our computation gives an exact integral representation for half-BPS correlators, without having to deal with a divergent generating series. Since we can express this generating function as a sum of residues with only one residue providing an exponentially large contribution, it is natural to expect that the saddle-point approximation gives the exact answer up to a simple one-loop determinant coming from the remaining residues. In fact the holographic computations of non-extremal correlators seem to agree with the exact computations obtained from explicit computations with the Schur basis \cite{Caputa:2012yj}. It would be nice to check whether this expectation holds by embedding the correlator into a supersymmetric observable where supersymmetric localization techniques can be used \cite{Pestun:2007rz}. For example, the connection between the coadjoint orbit integrals and Wilson loops via geometric quantization is well-known \cite{Witten:1988hf}.

From our saddle-point analysis, it seems that similar computation involving more generic mostly-symmetric Schur operators should proceed in the same way. More precisely, the HCIZ formula \eqref{HCIZ} gives in principle a sum over all possible pairings of the initial and final configurations of eigenvalues.  These can describe systems of more than one AdS giant graviton at different positions. Whenever the giants are well-separated we expect that the identity saddle dominates in a way that the computation reduces to a sum over the individual contributions of each giant. It would be more interesting to study set-ups where the positions of the branes coincide, in which case saddle points corresponding to permutations of equal eigenvalues are all relevant. It would be useful to understand the details in those cases before studying configurations of order $N^2$ stacked branes.

One surprising feature of these calculations is that the result is given by simple combinations of orthogonal polynomials in $\cosh{\rho_0}$ or $\cos{\theta_0}$ for AdS and sphere giants, respectively. This suggests that one might be able to compute these quantities by solving a wave equation with a non-trivial radial potential given by the presence of the branes. Understanding this connection would elucidate many of the physical aspects that are obscured in the present computations. One might expect that three-point functions with a spinning non-BPS single trace might be expressible as a spherical harmonic multiplied by a radial wavefunction. Also, since Legendre and Jacobi polynomials satisfy various recursive formulas, it might be possible to find non-trivial relations between BPS structure constants involving operators of different conformal dimensions in the large $N$ limit.

Another issue that needs attention is whether AdS giant gravitons can lead to integrable boundary states for the $\mathcal{N}=4$ SYM spin chain. Since many quantities associated with AdS giants can be obtained by analytic continuations from the sphere giant quantities, we expect that the answer to this question is negative, since non-maximal sphere giants do not appear to lead to integrable boundary states \cite{Chen:2019gsb}. However, there is new evidence that non-maximal giants do lead to integrable boundaries in the ABJM theory \cite{Yang:2021hrl}. Since there is no obvious reason for this qualitative difference, it would be useful to revisit some of these computations with new techniques.

\acknowledgments
AH would like to thank David Berenstein for collaboration on related topics. We also thank Shota Komatsu, Charlotte Kristjansen, and particularly Robert de Mello Koch for stimulating discussions, as well as the organizers of the KITP programs \textit{Confinement, Flux Tubes, and Large N} and \textit{Integrability in String, Field, and Condensed Matter Theory}. AH and WW were supported in part by funds from the University of California. A.H. was supported in part by the Department of Energy under grant DE-SC0011702. WW was supported in part by the U.S. Department of Energy under Grant No. DE-SC0023275.

\newpage

\appendix

\section{Details of Holographic Computation}

\subsection{Orbit Average of Diagonal Structure Constant} \label{app:diag_orbit}
We would like to compute the orbit average of the light operator insertion to get the diagonal structure constant
\bea
C_{\hat{\mathcal{S}}_{\Delta}\hat{\mathcal{S}}_{\Delta}\hat{\mathcal{O}}_L} = \int_{-\infty}^\infty d\tau_0 \int_0^{2\pi} \fr{d\phi_0}{2\pi} \mathcal{O}\[X^*_{\p_0,\t_0}\]
\eea
where
\bea
\mathcal{O}\[X^*_{\p_0,\t_0}\] &=& -\fr{1}{2\pi}\sqrt{L}(L+1) \sinh^2\r_0 \int_0^{\pi/2} d\wtd\c_1 \int_0^{2\pi} d\wtd\c_3 \sin\wtd\c_1 \cos\wtd\c_1 \\
&& \hspace{40mm} \times \(\fr{\cos\phi_0}{\cosh\r_0 \cosh \tau_0 - \cos\wtd\c_1\sin\wtd\c_3 \sinh\r_0}\)^L. \nn
\eea
To perform the average over moduli, it is useful to first Taylor expand the integrand in powers of $\lambda \cos\wtd\c_1\sin\wtd\c_3$, where $\lambda = \tanh\r_0 \cosh^{-1}\tau_0$, and perform the integrals over the $S^3$:
\bea
\int_0^{2\pi} d\wtd\c_3 \int_0^{\pi/2} d\wtd\c_1 \fr{\sin\wtd\c_1 \cos\wtd\c_1}{\(1 - \lambda \cos\wtd\c_1\sin\wtd\c_3 \)^{L+2}} &=& \int_0^{2\pi} d\wtd\c_3 \int_0^{\pi/2} d\wtd\c_1 \sin\wtd\c_1 \cos\wtd\c_1 \nn \\
&& \times \sum_{n=0}^\infty \fr{\Gamma(L + 2 + n)}{\Gamma(n + 1) \Gamma(L + 2)} \( \lambda \cos\wtd\c_1\sin\wtd\c_3\)^n \nn \\
&=& \int_0^{2\pi} d\wtd\c_3 \sum_{n=0}^\infty \fr{1}{n+2} \fr{\Gamma\(L + 2 + n\)}{\Gamma\(n + 1\) \Gamma\(L + 2\)} \( \lambda \sin\wtd\c_3\)^n \nn \\
&=& \fr{\pi}{\Gamma\(L + 2\)} \sum_{n=0}^\infty \fr{\Gamma\(L + 2 + 2n\)}{\Gamma\(n + 1\) \Gamma\(n+2\)} \(\fr{\lambda}{2}\)^{2n}. \nn
\eea
From this, we see that
\ba \label{eq:diag_app}
C_{\hat{\mathcal{S}}_{\Delta}\hat{\mathcal{S}}_{\Delta}\hat{\mathcal{O}}_L} &= -\fr{1}{2} \fr{\sqrt{L}}{\Gamma\(L + 1\)} \fr{\tanh^2\r_0}{\(\cosh\r_0 \)^{L}} \int_0^{2\pi} \fr{d\phi_0}{2\pi} \cos^L \phi_0 \\
& \hspace{-5mm} \times \sum_{n=0}^\infty \fr{\Gamma\(L + 2 + 2n\)}{\Gamma\(n + 1\) \Gamma\(n+2\)} \(\fr{\tanh\r_0}{2}\)^{2n} \int_{-\infty}^\infty \fr{d\tau_0}{\(\cosh\tau_0\)^{L+2+2n}}. \nn
\ea
The integrals over the $\phi_0$ and $\tau_0$ are then easily performed using
\bea
&& \int_0^{2\pi} \fr{d\phi_0}{2\pi} \cos^L\phi_0 = \fr{1^{L}+\(-1\)^{L}}{2^{L+1}} \fr{\G\(L+1\)}{\G\(\fr{L}{2}+1\)^2} \\
&& \int_{-\infty}^\infty \fr{d\tau_0}{\cosh^m\tau_0} = \fr{\sqrt{\pi}\,\G\(\fr{m}{2}\)}{\Gamma\(\fr{m+1}{2}\)},
\eea
and the result after summing over $n$ is
\bea \label{eq:diag_app_2}
C_{\hat{\mathcal{S}}_{\Delta}\hat{\mathcal{S}}_{\Delta}\hat{\mathcal{O}}_L}
&=& -\fr{1}{2} \(1 + (-1)^L\)\sqrt{L} \fr{\tanh^2\r_0}{\(\cosh\r_0 \)^{L}} \, {}_2F_1\(\fr{L}{2} + 1, \fr{L}{2} + 1, 2, \tanh^2\r_0\). \hspace{5mm}
\eea
This is equivalent to the gauge theory result \eqref{eq:diag_gauge} as we show in appendix \ref{app:matching}.

\subsection{Orbit Average of Off-diagonal Structure Constant} \label{app:offdiag_orbit}
The computation of the off-diagonal structure constants proceeds in a similar fashion, except now we must include the phases coming from the boundary wavefunctions
\bea
C_{\hat{\mathcal{S}}_{\Delta+k}\hat{\mathcal{S}}_{\Delta}\hat{\mathcal{O}}_L} = \int_{-\infty}^\infty d\tau_0 \int_0^{2\pi} \fr{d\phi_0}{2\pi} \mathcal{O}\[X^*_{\p_0,\t_0}\] e^{ik\phi_0} e^{-k\tau_0}.
\eea
What this amounts to is inserting the phases into \eqref{eq:diag_app}:
\ba \label{eq:offdiag_app}
C_{\hat{\mathcal{S}}_{\Delta+k}\hat{\mathcal{S}}_{\Delta}\hat{\mathcal{O}}_L} &= -\fr{1}{2} \fr{\sqrt{L}}{\Gamma\(L + 1\)} \fr{\sinh^2\r}{\(\cosh\r \)^{L+2}} \int_0^{2\pi} \fr{d\phi_0}{2\pi} \cos^L \phi_0 e^{ik\phi_0} \\
& \hspace{-5mm} \times \sum_{n=0}^\infty \fr{\Gamma\(L + 2 + 2n\)}{\Gamma\(n + 1\) \Gamma\(n+2\)} \(\fr{\tanh\r}{2}\)^{2n} \int_{-\infty}^\infty \fr{d\tau_0}{\(\cosh\tau_0\)^{L+2+2n}} e^{-k\tau_0}. \nn
\ea
It is easy to show that for $k \leq L$,
\bea
&&\int_{0}^{2\pi} \fr{d\phi_0}{2\pi} \cos^L \p_0 \,  e^{ik\p_0} = \fr{1^{L-k}+\(-1\)^{L-k}}{2^{L+1}} \fr{\G\(L+1\)}{\G\(\fr{L-k}{2}+1\)\G\(\fr{L+k}{2}+1\)} \label{eq:cos} \\
&& \int_{-\infty}^\infty \fr{d\tau_0}{\cosh^m\tau_0} \, e^{-k\tau_0}
= 2^{m-1} \frac{\Gamma\left(\frac{m-k}{2} \right)\Gamma\left(\frac{m+k}{2} \right)}{\Gamma\left(m\right)}.
\eea
Plugging these into \eqref{eq:offdiag_app} and performing the summation over $n$, we find
\ba
C_{\hat{\mathcal{S}}_{\Delta+k}\hat{\mathcal{S}}_{\Delta}\hat{\mathcal{O}}_L}
&= -\fr{1}{2}\(1^{L-k}+\(-1\)^{L-k}\) \sqrt{L} \\
& \hspace{5mm} \times \, \fr{\tanh^2\rho_0}{\cosh^L\rho_0} \, {}_2F_1\(1 + \frac{L-k}{2}, 1 + \frac{L+k}{2}, 2, \tanh^2 \r_0\). \nn
\ea
This can be shown to be equal to \eqref{eq:offdiag_gauge} by using the same set of identities used to turn the diagonal structure constant into Legendre polynomials in the next section.

For $k > L$, the integral \eqref{eq:cos} equals zero identically, leading to a vanishing structure constant.

\subsection{Matching to the gauge theory} \label{app:matching}
We now show that the holographic diagonal structure constant \eqref{eq:diag_app_2} is equivalent to the gauge theory result \eqref{eq:diag_gauge}. Setting $z = \tanh^2\r_0$, this amounts to showing
\bea \label{eq:matching}
P_{\frac{L}{2}}\left(\fr{1-z}{1+z}\right)-P_{\frac{L}{2}-1}\left(\fr{1-z}{1+z}\right) = \fr{L}{2} \, z (1-z)^{L/2} \, {}_2F_1\(1 + \fr{L}{2}, 1 + \fr{L}{2}, 2, z \).
\eea
Using the hypergeometric representation of the Legendre polynomials
\bea
P_{n}\left(\fr{1-z}{1+z}\right) = {}_2F_1\(-n, n+1, 1, \fr{z}{z-1} \)
\eea
the left-hand-side of \eqref{eq:matching} becomes
\bea
P_{\frac{L}{2}}\left(\cosh 2 \rho_0\right) - P_{\frac{L}{2}-1}\left(\cosh 2 \rho_0\right) &=& {}_2F_1\(-\fr{L}{2},\fr{L}{2} + 1, 1, \fr{z}{z-1} \) - {}_2F_1\(-\fr{L}{2}+1,\fr{L}{2}, 1, \fr{z}{z-1} \). \nn
\eea
Using the identity
\bea
{}_2F_1\(-n, a, b - n, \fr{z}{z - 1}\) = (1 - z)^a {}_2F_1\(a, b, b - n, z\)
\eea
on each term, we can write this as
\bea
P_{\frac{L}{2}}\left(\fr{1-z}{1+z}\right) - P_{\frac{L}{2}-1}\left(\fr{1-z}{1+z}\right) &=& (1 - z)^{\fr{L}{2}} \[(1 - z) \, {}_2F_1\(\fr{L}{2} + 1, \fr{L}{2}+1, 1, z\) - {}_2F_1\(\fr{L}{2}, \fr{L}{2}, 1, z\)\]. \nn
\eea
Finally, combining the identities
\ba
& {}_2F_1\(a, b - 1, 1, z\) + (z - 1) {}_2F_1\(a, b, 1, z\) + (a - 1) z {}_2F_1\(a, b, 2, z\) = 0 \nn \\
& {}_2F_1(a + 1, b, 1, z) - {}_2F_1\(a, b, 1, z\) = 0
\ea
we have \eqref{eq:matching} as desired.

\bibliographystyle{JHEP}
	\cleardoublepage
	
\renewcommand*{\bibname}{References}

\bibliography{references}
\end{document}